\documentclass[12pt,preprint]{aastex}

\newcommand{\mbfB}{\mathbf{B}}

\newcommand{\mbfv}{\mathbf{v}}
\newcommand{\mbfz}{\mathbf{z}}

\newcommand{\mbfx}{\mathbf{x}}
\newcommand{\mbfy}{\mathbf{y}}

\newcommand{\mbfnabla}{\mathbf{\nabla}}
\newcommand{\f}   {\frac}

\begin{document}

\title{Fast Dynamos in Weakly Ionized Gases}

\author{Ellen G. Zweibel\altaffilmark{1,2}}
\author{Fabian Heitsch\altaffilmark{3}}
\altaffiltext{1}{Depts. of Astronomy \& Physics, U. Wisconsin-Madison, 475 N Charter St, Madison,
                 WI 53706, U.S.A.}
\altaffiltext{2}{Center for Magnetic Self-Organization in Laboratory \& Astrophysical Plasmas}
\altaffiltext{3}{Dept. of Astronomy, U. Michigan, 500 Church St, Ann Arbor, MI 48109-1042, U.S.A.}

\lefthead{Zweibel \& Heitsch} 
\righthead{Fast Dynamos}

\begin{abstract}
The turnover of interstellar gas on $\sim 10^9$yr timescales argues for the 
continuous operation of a galactic dynamo. The conductivity of interstellar 
gas is so high that the dynamo must be ``fast" - i.e. the magnetic field must 
be amplified at a rate nearly independent of the magnetic diffusivity. Yet, 
all the fast dynamos so far known - and all direct numerical simulations of 
interstellar dynamos - yield magnetic power spectra that peak at the resistive 
scale, while galactic magnetic fields have substantial power on large scales. 
In this paper we show that in weakly ionized gas the limiting scale may be the 
ion-neutral decoupling scale, which although still small is many orders of magnitude 
larger than the resistive scale. 
\end{abstract}

\section{Introduction}\label{s:introduction}

Despite many theoretical and observational advances, our understanding of galactic magnetic fields
is still incomplete. Although there is evidence for magnetic fields in young galaxies (Kulsrud \& Zweibel 2008), it is likely that 
dynamo processes still operate 
continuously in galaxies today. Perhaps the most compelling argument for ongoing 
dynamo activity is that the turnover time of interstellar gas due to loss and replacement 
of material - $\sim$ 10$^9$ yr in the case of the Milky Way - is
much less than the ages of galaxies. As interstellar gas is added - whether by
infall from the IGM or through stellar mass loss - its magnetic field must be brought to the galactic value
so as to maintain the
field in a steady state. 

In contrast to the primordial situation at early times, when a magnetic field had to be
built up from nearly zero, it is likely that all the material 
on which galactic dynamos now act is at least somewhat magnetized. Under these conditions, the 
primary functions of the dynamo are to increase magnetic energy to the observed value, to 
generate and maintain a 
component of the field which is coherent over at least several kpc, as observed, and to transport magnetic flux into
the gas which has newly arrived, which especially if it is intergalactic in origin or was shed by stars, may be
under-magnetized. These processes are also necessary in theories such as that of Rees (1987), in which galactic magnetic fields
are seeded by random fields injected by many small scale sources. 

Dynamos are considered ``fast" if the rate at which the field is amplified tends 
to a finite limit as the magnetic diffusivity $\eta$ approaches zero 
(Childress \& Gilbert 1995). It may seem paradoxical that magnetic diffusivity, 
which removes energy from the magnetic field, is required for dynamos at all. But 
without diffusion the magnetic topology is fixed, and this places strong 
constraints on magnetic field amplification (Moffat 1978). The Ohmic diffusion 
time in galaxies is so much longer than any dynamical time that it seems galactic
dynamos must be fast.

Fast dynamo theory has followed two approaches. One involves computation and 
analysis of the action of prescribed flows on magnetic fields
under nearly non-diffusive conditions (e.g. Galloway \& Proctor 1992, Ott 1998, Gilbert 2002, Couvoisier, Hughes, \&
Tobias 2006). These so-called kinematic studies 
have demonstrated the role of chaotic flow in magnetic field amplification and 
elucidated important relationships between the topological properties of the 
flows and their dynamo properties (Klapper \& Young 1995). The modification of some of these flows by 
magnetic forces has also been considered, and has provided some insight into 
how dynamos might saturate (Cattaneo, Hughes, \& Kim 1996,
Tanner \& Hughes 2003, Cameron \& Galloway 2005).

The second approach to fast dynamos involves analysis and direct numerical simulations of driven 
magnetohydrodynamic (MHD) turbulence, again with the lowest magnetic diffusivity  
$\lambda$ possible. In particular, $\lambda$ is chosen to be less than the viscous 
diffusivity $\nu$ (i.e. the large Prandtl number case). These models are
self-consistent in the sense that magnetic forces are fully included in the 
dynamics, and theory and simulation are in good agreement on how the field evolves (Schekochihin et al. 2004).

Both approaches yield magnetic fields which are 
dominated by structure on the resistive scale, which is far smaller than any other 
characteristic scale in the interstellar medium. This follows from the
fact that in the absence of magnetic diffusion, any divergence free flow which 
amplifies the field lengthens the field lines in the same proportion. Therefore, 
as the field is amplified it becomes folded or tangled. The large random field is in stark contrast 
to the observed structure of galactic magnetic fields,
which display  considerable long range order, and thus presents a serious challenge 
to dynamo theory. In particular, 
the magnetic field is amplified on scales far below 
the minimum velocity scale, 
whether this scale is prescribed (the kinematic case) 
or created by viscous effects (the full turbulent simulation).

Most of the mass, and a considerable part of the volume of the interstellar 
medium, is weakly ionized, with the plasma and neutral fluids coupled by 
collisions and by ionization and recombination. On long lengthscales 
and timescales the medium behaves like a single conducting fluid, but on short 
lengthscales and timescales, the two fluids decouple. For present interstellar 
medium parameters, the magnetic field is in approximate energy equipartition with the
bulk fluid and dominates the plasma component. The question therefore arises whether 
a dynamo in a weakly ionized medium could be quenched on scales below the 
plasma-neutral decoupling scale while continuing to amplify the field on larger 
scales. Based on analysis and computations presented in this paper, the answer 
seems to be that it can. 

This is far from the first study of the effects of partial ionization on dynamos. There have been a number of studies of large
scale dynamos in weakly ionized gases which show that the nonlinearity introduced by ion-neutral drifts has important
consequences for mean field theory (Zweibel 1988,  Proctor \& Zweibel 1992)  and for the general problem of how large scale fields
are amplified in a medium with small scale turbulence (Kulsrud \&
Anderson 1992, Subramanian 1997, 1999, Brandenburg \& Subramanian 2000). These
studies all used the strong coupling approximation (Shu 1983), according to which the plasma drift relative to the
neutrals is found by balancing the Lorentz force by ion-neutral drag. This approximation breaks down at small scales.
In this work, we treat
the plasma and neutrals as two frictionally coupled fluids, which is more appropriate at the small scales, and also more general.

In order to see how ion-neutral friction might affect the dynamo, consider the equation for magnetic energy, which in a domain $V$
with periodic or infinitely distant boundaries can be written as
\begin{equation}\label{e:adenergy}
\f{\partial}{\partial t}\int_V\f{B^2}{8\pi}d^3x = -\int_V\left[\mbfv_i\cdot\f{(\mbfnabla\times\mbfB)\times\mbfB}{4\pi} +
\lambda\f{\vert\mbfnabla\times\mbfB\vert^2}{4\pi}\right]d^3x,
\end{equation}
where $\mbfv_i$ is the plasma velocity and $\lambda$ is the Ohmic diffusivity. 
The first term in the integral on the right hand side represents the work down by the field on the plasma; the second term is
resistive dissipation. 

The plasma velocity can always be expressed in terms of the neutral velocity $\mbfv_n$ and the drift
velocity $\mbfv_D\equiv \mbfv_i-\mbfv_n$; $\mbfv_i=\mbfv_v+\mbfv_D$. If the ionization fraction is low, $\mbfv_n$, in turn, is
very nearly the center of mass velocity $\mbfv$. In the strong coupling approximation, $\mbfv_D$ is
\begin{equation}\label{e:sca}
\mbfv_D=\f{(\mbfnabla\times\mbfB)\times\mbfB}{4\pi\gamma\rho_i\rho_n},
\end{equation}
where $\gamma\equiv\left<\sigma v\right>_{in}/(m_i+m_n)$
is the ion-neutral friction coefficient.
Replacing $\mbfv_i$ by $\mbfv+\mbfv_D$ in equation~(\ref{e:adenergy}) and using equation~(\ref{e:sca}), we see that when the strong
coupling approximation holds, the magnetic energy equation is
\begin{equation}\label{e:adenergy2}
\f{\partial}{\partial t}\int_V\f{B^2}{8\pi}d^3x = -\int_V\left[\mbfv\cdot\f{(\mbfnabla\times\mbfB)\times\mbfB}{4\pi} +
\rho_i\rho_n\gamma v_D^2 +\lambda\f{\vert\mbfnabla\times\mbfB\vert^2}{4\pi}\right]d^3x.
\end{equation}
Equation (\ref{e:adenergy2}) shows, as expected, that ion-neutral friction is a dissipative effect. The work term, however, can
have either sign, and it is possible that in the presence of ambipolar drift, the magnetic field and flow are modified so as
to increase the rate at which energy flows to the field. Equation (\ref{e:adenergy}), on the
other hand, is valid whether the strong coupling
approximation holds or not, is closer to the magnetic energy equation for a plasma,
and does not attempt to split the inductive and dissipative  effects.

In \S{\ref{s:setup}} we introduce the basic setup. In \S{\ref{s:results}} we show 
how plasma-neutral friction can allow the field to saturate at the small scales 
while still being amplified at the large scales. Section {\ref{s:summary}} is a summary.

\section{Formulation}\label{s:setup}
\subsection{Important timescales and lengthscales}

In a turbulent cascade, the velocity at scale $l$, $v_l$, typically decreases with decreasing $l$, but slowly enough that the
turnover time $\tau_l\equiv l/v_l$, decreases with $l$ as well. For example, in Kolmogorov turbulence, or
in the magnetohydrodynamic
turbulence model of Sridhar \& Goldreich (1994) and
Goldreich \& Sridhar (1995, 1997), $v_l$ is related to $l$ and the scale $l_d$ at which the turbulence is
driven by $v_l\sim v_d(l/l_d)^{1/3}$. The turnover time $\tau_l$ is then $(l_d/v_d)(l/l_d)^{2/3}$. The cascade terminates at the scale
$l_K$ at which the dissipation rate becomes faster than the turnover rate.

In diffuse, weakly ionized interstellar gas, $\tau_K$, the turnover time at $l_K$, is comparable to the neutral-ion collision time
$\tau_{ni}$, and much longer than the ion-neutral collision time $\tau_{in}$. Only for motions on timescales less than $\tau_{in}$ is
friction with the neutrals unimportant for plasma dynamics. Resistive effects are typically unimportant for motions with turnover times
$\tau_{in}$, but become important on much smaller scales. The ion viscous scale is probably terminates the ion flow on scales much larger
than the resistive scale (the suppression of viscous stresses perpendicular to $\mbfB$ even for a weak magnetic field makes the actual
value difficult to assess), 
and the 

The computational resources available to us preclude modeling all these scales in a realistic 
manner. Our main focus is on separating the
ion-neutral decoupling scale and the resistive scale. As a consequence, instead of 
assigning the neutral motions a third, still
larger scale, we allow the neutral flow to take place on 
the ion-neutral decoupling scale.  Furthermore, we rely entirely on numerical viscosity, which allows the
flow in the ions to extend below the resistive scale. In reality, ion viscosity probably terminates the ion flow on scales much larger
than the resistive scale (the suppression of viscous stresses perpendicular to $\mbfB$ even for a weak magnetic field makes the actual
value difficult to assess), but again we are unable to achieve true scale separation (or implement anisotropic viscosity).
 
The consequences of these 
aspects of our calculations are discussed in the
next two sections.

\subsection{Case study}
We assume that the neutral velocity $\mbfv_n$ is of a form which would give fast dynamo action 
if it were the plasma velocity $\mbfv_i$. On timescales much longer than the ion-neutral collision time $(\rho_n\gamma_{in})^{-1}$, 
the plasma should move with the neutrals - $\mbfv_i\sim\mbfv_n$ - and the magnetic field should 
grow at the fast dynamo rate, at least while Lorentz forces are unimportant. On short lengthscales, 
inertial and Lorentz forces compete with ion-neutral friction to drive $\mbfv_i$ away from 
$\mbfv_n$. We investigate the dynamo properties of the resulting plasma flow. The setup is somewhat similar to the two-fluid study of
Kim (1997), although that work focused on diffusion of a large scale field, and the computations were two dimensional and incompressible.

We simplify the problem by assuming the ionization  and recombination timescales are much 
shorter than the advective timescales, so that ionization equilibrium holds. This assumption
allows us to bypass the plasma continuity equation, but it 
is unrealistic in the diffuse gas
which is the primary application for this study [although it is much better for molecular gas (see Table 3 of Heitsch \& Zweibel
2003a)].
The neutral velocity we choose is divergence free, so by taking the neutral 
density $\rho_n$ to be initially uniform, we force it, and the plasma density $\rho_p\sim\rho_i$, 
to remain so. Therefore, there are no thermal pressure gradient forces in the problem. Under these 
conditions, the momentum equation for the plasma is
\begin{equation}\label{e:ionm1}
\rho_i\left(\frac{\partial}{\partial t}+\mbfv_i\cdot\mbfnabla\right)\mbfv_i=
\f{\left(\mbfnabla\times\mbfB\right)\times\mbfB}{4\pi}
-\gamma\rho_i\rho_n\left(\mbfv_i-\mbfv_{n}\right),
\end{equation}
where 
$\mbfv_n$ is to be prescribed.
Equation~(\ref{e:ionm1}) is to be solved together with the magnetic induction equation
\begin{equation}\label{e:induction}
\f{\partial\mbfB}{\partial t}=\mbfnabla\times(\mbfv_i\times\mbfB)+\lambda\nabla^2\mbfB.
\end{equation}

For $\mbfv_n$ we choose the 2.5D
flow shown by Galloway \& Proctor (1992) to be a fast dynamo. This flow, which we will
refer to as $\mbfv_{GP}$, can be written as
\begin{equation}\label{e:gpflow}
\mbfv_n=\hat{x}2\pi k_{GP}\psi+\hat{y}\frac{\partial\psi}{\partial z}-\hat{z}\frac{\partial\psi}{\partial y},
\end{equation}
where the stream function $\psi$ is
\begin{eqnarray}\label{e:psi}
\psi =\sqrt{\frac{3}{2}}\frac{V_0}{2\pi k_{GP}}\Big(& &\sin{[2\pi( k_{GP}z+\epsilon\sin{2\pi\omega t})]}\nonumber\\
     &+&\cos{[2\pi(k_{GP}y+\epsilon\cos{2\pi\omega t})]}\Big),
\end{eqnarray}
and $\epsilon$ is a constant. This flow has periodic, cellular structure on a single spatial scale, 
$k^{-1}$, and temporal structure at frequency
$\omega$ and all of its harmonics. This can readily be seen by rewriting equation~(\ref{e:psi}) in the form
\begin{eqnarray}\label{e:psiseries}
\psi =\sum_{n=0}^{\infty}\psi_n
 \Big(& &\sin{[2\pi (k_{GP}z+n\omega t)]} \nonumber\\ 
      &+&\cos{[2\pi (k_{GP}y+n(\omega t + \frac{1}{4})]}\Big),
\end{eqnarray}
where 
\begin{equation}\label{e:psin}
\psi_n\equiv \sqrt{\frac{3}{2}}\frac{V_0}{2\pi k_{GP}}J_n(2\pi\epsilon)
\end{equation}
and the $J_n$ are ordinary Bessell functions of the first kind. However, bearing in mind that 
$J_n(x)\sim (x/2)^n/n!$ for $x<n$, and that $x=\pi/2$ in our problem, we see that only the first 
few harmonics of the series are important in the flow. The cell pattern drifts back and 
forth with temporal frequency $\omega$ and amplitude $\epsilon$. The flow  has chaotic regions 
along the cell edges. In this paper, we choose units of  length and time such that 
$k_{GP}= V_0 = \omega \equiv 1$.
The eddy turnover time and the period over
which the pattern oscillates are then of the same order. We set $\epsilon\equiv 0.25$, near the value 
at which chaos is maximized (Brummell, Cattaneo, \& Tobias (2001)). The ranges of $y$ and $z$ are 
$-1\le y\le 1$, $-1\le z\le 1$, so only one GP cell fits into the domain. We choose $\lambda$ 
such that $\lambda k_{GP}^2$, the resistive decay rate at the GP scale, is between $10^{-2}$ and 
$10^{-4}$, and the friction coefficient, $\gamma_{in}\rho_n$, is between $0.03$ and $0.60$. We set 
$\rho_i=1$. The magnetic field is initialized by choosing a vector potential which is a Fourier 
series in modes of the computational domain with random amplitudes and random
phases selected from uniform distributions.

The numerical scheme, Proteus, is based on the conservative gas-kinetic flux splitting method 
introduced by Xu (1999) and Tang \& Xu (2000), and is second order in time and space. 
Resistivity is included through dissipative fluxes (Heitsch et al. 2004, 2007), but 
viscosity is not (there is a small amount of numerical viscosity).  The frictional force 
is implemented through a drag term (Heitsch et al. 2004).

\section{Results}\label{s:results}

\begin{figure*}
  \includegraphics[width=\textwidth]{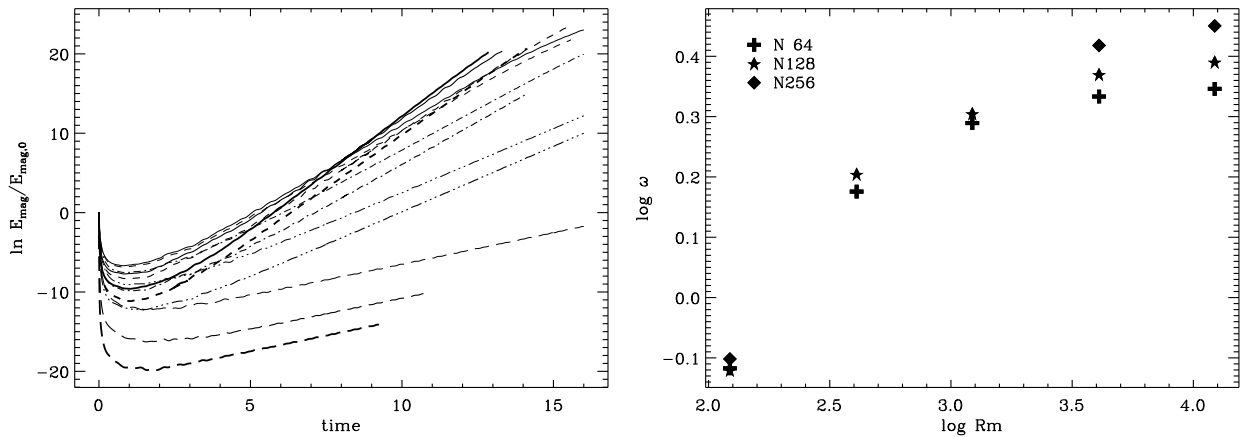}
  \caption{\label{f:allki}{\em Left:} Natural logarithm of the normalized magnetic energy against
           time for all kinematic models (series KI), at linear resolutions of $N=64$ (thin lines),
           $128$ (medium lines), and $256$ (thick lines). The higher $\lambda$, the stronger the
           initial decay of the (noise-like) seed field. {\em Right:} (Exponential) growth rates
           derived from fits to the curves in the left panel, for all resolutions and magnetic Reynolds numbers.}
\end{figure*}

The complete set of models we considered is summarized in Table 1.
We first verified that our computational scheme recovers the previously known 
dynamo behavior of the GP flow, and assessed numerical convergence. 
The left panel of Figure~\ref{f:allki} shows the  
magnetic energy  against time for three resolutions and five diffusivities. 
Once the short wavelength magnetic power present in the random
initial conditions has 
decayed, the magnetic energy grows at an exponential rate. The growth 
rates are plotted in the right panel of Figure~\ref{f:allki}. With our choice of units, $R_m=\lambda^{-1}$. 
While some flattening of the curves is apparent, the growth rate has evidently not
converged to the $R_m\rightarrow\infty$ limit. Nevertheless, the evident lack of convergence 
at the largest $R_m$ and highest numerical resolution practical for us shows that it would not 
be meaningful to set $\lambda$ to any value smaller than what we used here. This series of models 
will be referred to as KI (for model keys, see Table~\ref{t:models}). All the models discussed subsequently have $R_m=10^{4}$.

\clearpage
\clearpage

\begin{deluxetable}{l|cc}
  \tablewidth{0pt}
  \tablecaption{Key to model names\label{t:models}}
  \tablehead{\colhead{Name}&\colhead{Mnemonic}
             &\colhead{Physics}}
  \startdata
  KI  & Kinematic, Ions     & single-fluid, GP flow in ions\\
  KN  & Kinematic, Neutrals & two-fluid, GP flow in neutrals\\
  DL  & Dynamic, Lorentz    & two-fluid, GP flow in neutrals, Lorentz force\\
  DR  & Dynamic, Reynolds   & two-fluid, GP flow in neutrals, Reynolds Stress\\
  DA  & Dynamic, All        & DL combined with DR
  \enddata
\end{deluxetable}
\clearpage

\begin{figure*}
  \includegraphics[width=\textwidth]{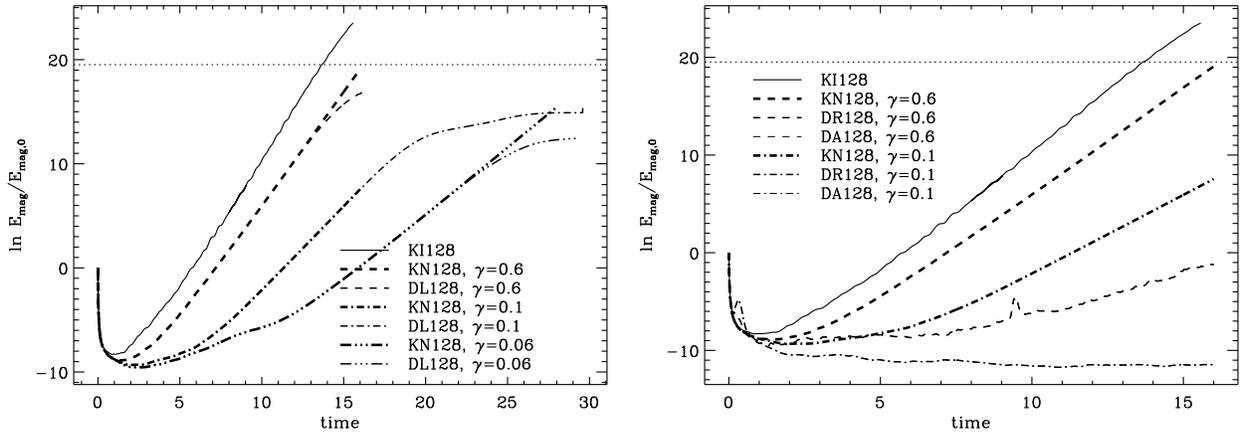}
  \caption{\label{f:alltimes}Natural logarithm of the normalized magnetic energy against
           time for models at linear resolution $N=128$. See Table~\ref{t:models} for 
           keys to the model names. Back-reaction via Lorentz-forces let the field amplification
           saturate (models KN vs DL), while the inclusion of the Reynolds stress suppresses the
           dynamo (models KN vs DR and DA). Note that there is no difference in the energy evolution
           between models DR and DA because the magnetic field simply stays too weak for the 
           Lorentz force to be of importance. The dashed horizontal line denotes energy equipartition
           between magnetic energy and the nominal kinetic energy in the GP flow.}
\end{figure*}

We next set $\mbfv_n=\mbfv_{GP}$ and solved equation~(\ref{e:ionm1}) with the Lorentz force and 
Reynolds stress terms dropped. These models, which we denote by KN, demonstrate the interplay 
between inertia and friction in determining the ion flow. Using equation~(\ref{e:psiseries}), 
equation~(\ref{e:ionm1}) can then be reduced to a series of equations for the $j$th Fourier harmonic of $\mbfv_i$
\begin{eqnarray}\label{e:vidot}
\left(\f{\partial}{\partial t}+\rho_n\gamma_{in}\right)\mbfv_{ij}
&=&\rho_n\gamma_{in}\psi_j\left(\hat\mbfx+\hat\mbfy\f{\partial}{\partial z}-
\hat\mbfz\f{\partial}{\partial y}\right)\nonumber \\
&\times& \Big(\sin{[2\pi( k_{GP}z+j\omega t)]}\nonumber\\ 
&+&\cos{[2\pi(k_{GP}y+j[\omega t+\frac{1}{4}])]}\Big),
\end{eqnarray}
which can be solved analytically. For the
parameters chosen here, $\rho_n\gamma_{in}\ll j\omega$ for $j>0$, and $\mbfv_{ij}$ is nearly $\pi/2$ out of phase with 
the frictional forcing term, and reduced in amplitude by a factor of 
$\rho_n\gamma_{in}/j\omega$. Thus, inertial effects force the plasma
away from the GP flow (this is an artifact of the overlap in our models between the turnover rate of the neutral eddies and the ion-neutral
collision rate; in the interstellar medium the latter is much larger). 
However, even though the plasma flow departs significantly from the GP flow, it does amplify
the magnetic field. This is shown in Figure~\ref{f:alltimes}.
The amplification rate decreases with decreasing $\rho_n\gamma_{in}$. This is caused
primarily by the reduced amplitude of the ion flow, but also by its structure, which as $\rho_n\gamma_{in}$ decreases departs more and
more from $\mbfv_{GP}$.
The dissipative nature of ion-neutral friction may also be playing a role here; see equation~(\ref{e:adenergy2}) - although equation~(\ref{e:sca}) is not completely accurate in this case.

It can be seen from equation
(\ref{e:induction}) that magnetic fluctuations at wave number $k$ interacting 
with flow at the GP wave number $k_{GP}$ are coupled to fields
at wave number $k\pm k_{GP}$. This sets off a magnetic cascade in wavenumber, and therefore, both
the GP flow and the modified flow present in the KI models drive magnetic field at all spatial scales. 
\clearpage
\clearpage

The Reynolds stress $\mbfv_i\cdot\mbfnabla\mbfv_i$, 
because of its nonlinearity, drives higher spatial harmonics of the GP wave number. In our parameter regime, it is of
the same order as the inertial term,  at least if $v_i$ is of the same magnitude as $v_{GP}$. 
Because in our calculation the resistivity exceeds
the (numerical) viscosity, the ion flow extends to smaller scales than the magnetic field. 
It is known that dynamo activity is often
suppressed in these so-called low magnetic Prandtl number situations unless the 
magnetic Reynolds number is large (Boldyrev
\& Cattaneo 2004, Schekochihin et al 2007). Because the ion density remains constant, the compressibility of the medium
is exaggerated; suppression of the dynamo has also been associated with compressibility effects (Haugen, Brandenburg, \& Mee,
2004). For both these reasons, it is perhaps unsurprising that including the Reynolds stress obliterates the dynamo; see
the DR models in Figure~\ref{f:alltimes}. 

\begin{figure*}
  \begin{center}
  \includegraphics[width=0.7\textwidth]{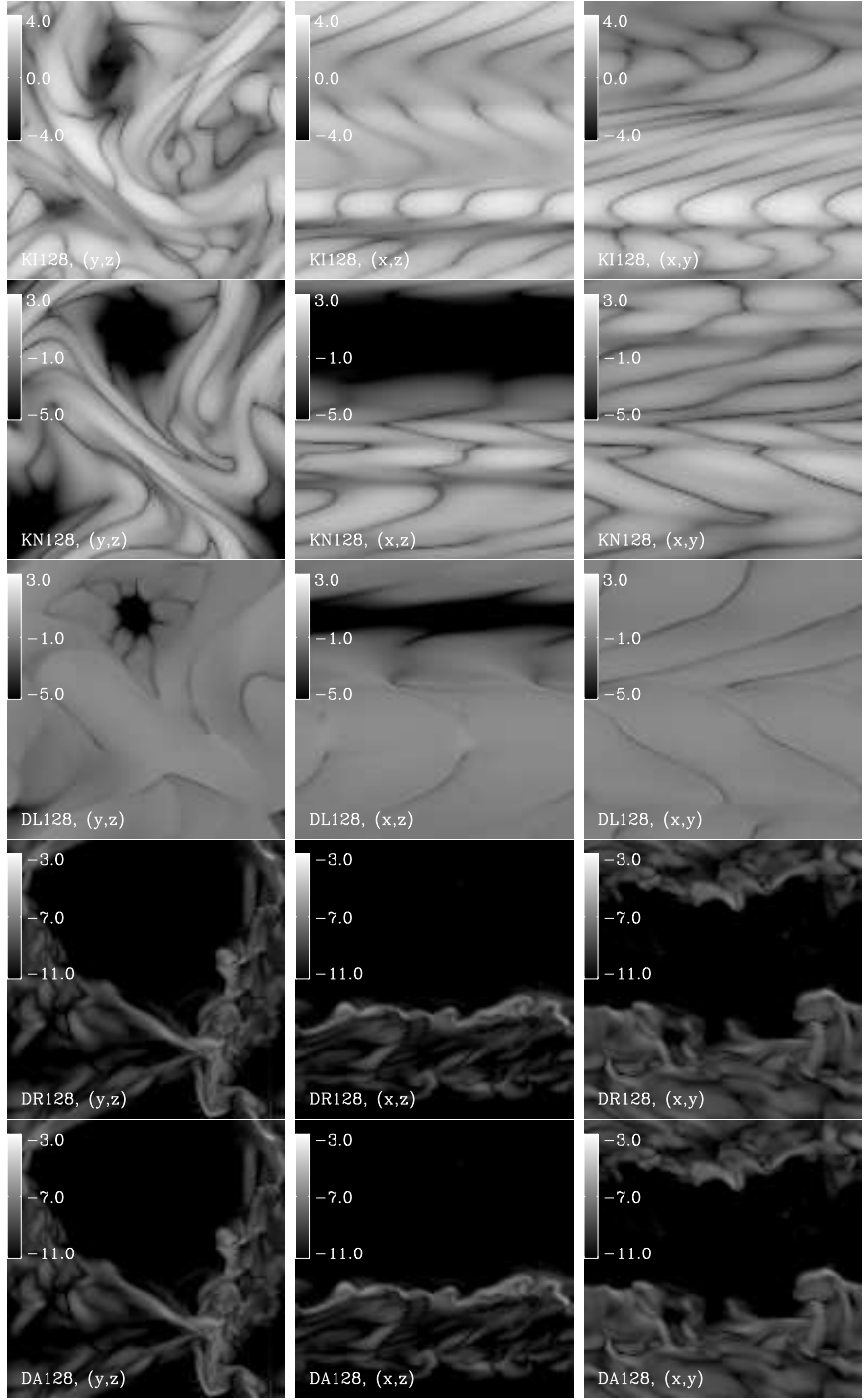}
  \end{center}
  \caption{\label{f:midplanes}Logarithm of magnetic energy density at midplane, in planes 
           $(y,z)$ (left), $(x,z)$ (center) and $(x,y)$ (right). Models from top to bottom:
           KI, KN, DL, DR and DA. For model names, see Table~\ref{t:models}.}
\end{figure*}

We also considered the effect of dropping the Reynolds stress term but including the Lorentz 
force (models DL). Because it is nonlinear in $B$, and because magnetic power is driven at all scales, the
Lorentz force too drives higher spatial harmonics in the flow. 
The feedback from the magnetic 
field eventually saturates the dynamo (Fig.~\ref{f:alltimes}). At the time of saturation, 
the volume integrated plasma kinetic energy is
about 100 times larger than the magnetic energy. However, magnetic forces along the cell 
walls, where the field is concentrated (see
Figure \ref{f:midplanes}), are locally strong, and inhibit the exponentially fast 
stretching needed for further amplification. This is
consistent with the results of Cattaneo, Hughes, \& Kim  (1996), who showed that magnetic forces 
suppress chaos - characterized by exponential stretching - in the Galloway-Proctor flow even when
the field is too weak to modify the flow kinetic energy. 

Models with the most complete physics, namely both Reynolds stress and Lorentz forces in addition 
to inertia and ion-neutral friction, fail to be dynamos (Fig.~\ref{f:alltimes}; models DA). This is to be expected,
since the initial magnetic field is too weak to affect the dynamics, and nonlinear 
advection alone suppresses the dynamo.

It is revealing to look at the structure of the magnetic field as well as its growth rate. 
Figure~\ref{f:midplanes}
shows the logarithm of the magnetic energy density 
from three views (along the $x$, $y$, and 
$z$ axes) for the KI, KN, DL, DR, and DA models, at $t=20$. The top set of panels represent the 
canonical kinematic GP dynamo (KI). The dark lines, where the magnetic field is very small, 
demonstrate that it is folded on scales much smaller than the flow scale, 
as is characteristic of dynamos with large $P_m$. 
The second row of 
panels shows the field when the plasma flow is determined by friction and inertia, but not Lorentz 
forces or Reynolds stresses (KN). It is weaker than in models KI, but still spatially intermittent 
and folded. If these folds were true magnetic nulls, they could be sites of fast magnetic 
reconnection, but even a small amount of magnetic shear is enough to quench this process 
(Heitsch \& Zweibel 2003a,b).

\clearpage
\clearpage

More dramatic changes come with the inclusion of the Lorentz force, but not the Reynolds 
stress (third row of Figure 3, models DL). Although Figure~\ref{f:alltimes} shows that the magnetic field is 
still some way from saturation, and the magnetic energy is indistinguishable from that of the 
kinematic models,the spatial intermittency and amount of folding are markedly reduced.

The final two rows, models DR and DA, show how the small scale flow generated by nonlinear 
advection affects the magnetic field. The ordered orientation associated with the strong 
shear in the GP flow is essentially gone. Since there is little dynamo action, magnetic 
feedback is unimportant, and the magnetic fields in the two cases are virtually identical. 
Although there has been little amplification of the field as a whole, there are thin 
filaments where it is locally strong. 
These structures are associated with strong compression (negative divergence in the ion
velocity) and with strong vorticity.

\begin{figure}
  \begin{center}
  \includegraphics[width=\columnwidth]{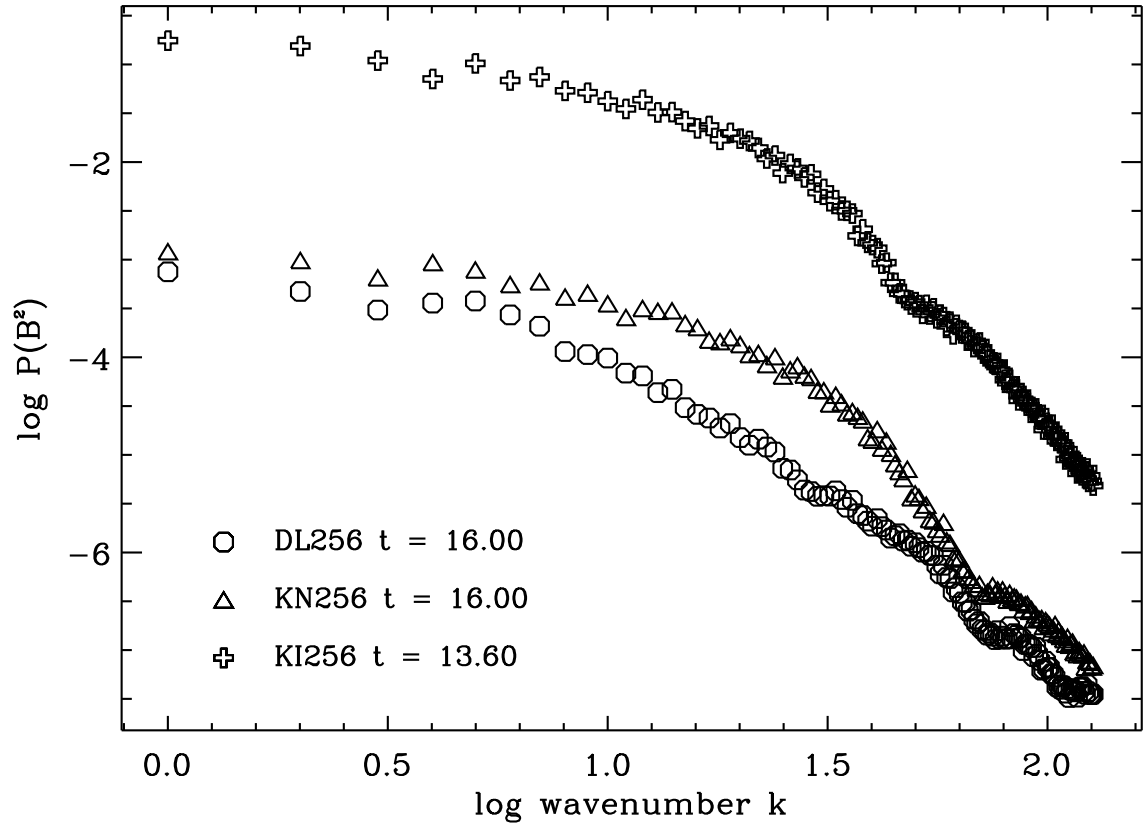}
  \end{center}
  \caption{\label{f:spectra}Power spectra of the magnetic energy for models KI, KN and DL at
           a linear resolution of $N=256$, and at system time $t=16$, i.e. at a point when
           the energy evolution of models KN and DL has separated due to saturation 
           (see Fig.~\ref{f:alltimes}).}
\end{figure}

Magnetic energy spectra for the KI, KN, and DL models at a linear resolution of $N=256$ are shown in 
Figure~\ref{f:spectra}. Although there is a 
slight reduction of power at the smallest scales it is not as important as the overall diminished 
amplitude and it is clear that power spectra without phase information give an incomplete picture of
the structure of the magnetic field.

\clearpage
\clearpage

\section{
Discussion and
Summary}\label{s:summary}

Galactic magnetic fields show considerable long range order, while computations
of small scale turbulent dynamos in highly conducting plasmas have shown instead that magnetic energy is concentrated at the
resistive scale. This suggests that there must be some feature of the interstellar dynamo which suppresses growth of the small 
scale field, 
and that there exist mechanisms for transferring energy to scales larger than the turbulent injection scale.
 
In this paper we investigated the role of ion-neutral friction in eliminating the smallest scales. We 
solved a simple model problem in which the neutral flow was known to be a dynamo, and was coupled to the plasma
through frictional drag. The plasma flow was determined by friction, inertia, and Lorentz forces. Ionization equilibrium 
was assumed. 
We used a spatially constant Ohmic resistivity, and we relied on viscous dissipation 
at the resolution scale. 

This study is, as far as we know, the first dynamo model in which the plasma flow is calculated from the full equation of motion,
rather than assuming the strong coupling approximation (Shu 1983). The strong coupling approximation breaks down at small scales,
and leads to an equation for magnetic energy growth which differs somewhat from the standard equation in a plasma [compare equations~(\ref{e:adenergy}) and (\ref{e:adenergy2})]. However, our
models differ from the interstellar medium in several significant respects, due to the prohibitively large computational resources
required for a realistic simulation. In the models, the 
neutral forcing occurs on
timescales comparable to the ion-neutral collision time, whereas in the diffuse,
weakly ionized interstellar medium the shortest timescale in neutral turbulence
is probably 2-3 orders of magnitude longer. This introduces an artificially large difference between $\mbfv_i$ and $\mbfv_n$.  By taking 
the magnetic diffusivity larger than the viscous diffusivity (small magnetic Prandtl number $P_m$), we reverse the interstellar ordering
($P_m\gg 1$). By assuming ionization equilibrium, an isothermal equation of state,
 and a uniform density of neutrals, we preclude the possibility of plasma pressure gradients. And finally, the resistive scale is only about
two orders of magnitude less than the dynamical scale, instead of 5-6, as in the ISM.

Our work does not address the growth of the field at the largest scales. Whether ambipolar diffusion acting on small scale
turbulence can drive a large scale field, either by quasilinear (Zweibel 1988, Proctor \& Zweibel 1992) or nonlinear
(Subrahmanian 1997, 1999, Brandenburg \& Subrahmanian 2000) effects cannot be probed by our study due to the large range of
scales that would be required, and also, perhaps, because the neutral velocity is already a dynamo.

Our main results are as follows. First, friction successfully imprints a dynamo flow on the ions, despite the importance of inertial terms
[models KN; Fig. (\ref{f:alltimes})]. That is, computing $\mbfv_i$ from equation~(\ref{e:vidot}) leads to a flow which amplifies a weak magnetic field. Second,
adding nonlinear advection ($\mbfv_i\cdot\mbfnabla\mbfv_i$) to equation~(\ref{e:vidot}) results in an ion flow which does {\textbf{not}}
amplify the magnetic field, except in thin regions with large vorticity or negative divergence [models DR; Fig. (\ref{f:midplanes})].
This may be a manifestation of the difficulty of driving a dynamo when $P_m\ll 1$, 
or in a highly compressible medium. It would not necessarily carry over
to the diffuse ISM, in which $P_m$ is large{\footnote{The viscosity tensor in interstellar plasma is highly anisotropic due to the
large ratio of the ion gyrofrequency to the Coulomb collision rate. Therefore, 
viscous transport across the magnetic field is strongly suppressed (Braginskii 1965), making the actual value of $P_m$ somewhat
unclear.}} and the recombination time is longer than the eddy turnover time.
 Third, when the Lorentz force is added, but the Reynolds stress is dropped, the dynamo saturates when the magnetic energy
is about two orders of magnitude less than the plasma kinetic energy [models DL; Fig. (\ref{f:alltimes})]. 
According to conventional wisdom, the magnetic energy
saturates only close to equipartition; the difference here
is due to the spatial intermittency of the magnetic field, which allows Lorentz forces to become strong just where the
field is being amplified. Finally, adding the Lorentz force suppresses small scale structure (Fig. \ref{f:midplanes}).

The saturation reported here for models without nonlinear advection (models DL) has some 
features in common with the study by Tanner \& Hughes (2003), although these authors 
considered fully ionized systems. They studied a superposition of the GP flow (which they 
call the CP, or Circularly Polarized flow) and a steady flow known to be a slow dynamo, 
restricted the dynamics by omitting the Reynolds stress term and averaged the Lorentz force
over what in our case would be the $x$ direction. They found that when the GP flow dominates, 
saturation occurs through suppression of exponential field line stretching by Lorentz forces, 
rather than by enhanced dissipation brought about by an increase in small scale structure.

Taken at face value, our computations suggest that in weakly ionized gases, the efficiency of 
dynamos is reduced at scales below the neutral-ion coupling length. The
main point is that saturation of the field 
at small scales is determined by the kinetic energy
in the plasma at these scales rather than in the bulk medium. 
The inertia of the neutrals contributes, and presumably increases the saturation level, only
for motions with turnover time $\tau_l$ greater than the neutral-ion collision time $\tau_{ni}\equiv (\rho_i\gamma_{in})^{-1}$. If we take
$n_i\sim 2\times 10^{-3}$ cm$^{-3}$, representative of both cold and warm neutral interstellar gas, then $\tau_{ni}\sim 3\times 10^{11}$s.
Taking the expressions for
Kolmogorov turbulence given in \S 2.1, and assuming the  driving scale and velocity are 10 pc and 10 km s$^{-1}$, 
respectively, we find $\tau_l\sim
\tau_{ni}$ at $l\sim 10^{-2}$pc. The equipartition magnetic field strength at this scale relative to the ions is only $\sim .02\mu G$, or about
1\% of the Galactic value. Thus, saturation of the dynamo below the neutral-ion decoupling scale could prevent amplification of the field
at very small scales. 

At early times, when the magnetic field might have been several orders of magnitude or more weaker
than it is today, the decoupling effect would have been less important. Furthermore, the 
interstellar medium was probably almost fully ionized until metallic coolants had mixed into it. 
Thus, the mechanisms addressed here are probably more relevant to the maintenance of galactic 
magnetic fields than to their ultimate origin.

\acknowledgements 
We are happy to acknowledge supports by NSF Grants AST-0507367 and 
PHY-0215581 to the University of Wisconsin. Parts of the computations were performed at
the National Center for Supercomputing Applications under projects DAC AST-040026 and 
MRAC AST-060031. We are grateful to an anonymous referee for useful suggestions.


\begin{thebibliography}{}
\bibitem[Boldyrev \& Cattaneo(2004)]{517}
        Boldyrev, S. \& Cattaneo, F. 2004, \prl, 92, 144501
\bibitem[Brandenburg \& Subramanian(2000)]{}
	Brandenburg, A. \& Subramanian, K. 2000, \aa, 361, L33
\bibitem[Brummell, Cattaneo \& Tobias(2001)]{BCT2001}
        Brummell, N.~H., Cattaneo, F., Tobias, S.~M.\ 2001,
        {\em Fl. Dyn. Res.}, 28, 237
\bibitem[Cameron \& Galloway(2005)]{522}
        Cameron, R. \& Galloway, D.\ 2005, \mnras, 358, 1025
\bibitem[Cattaneo, Hughes, \& Kim(1996)]{524}
	Cattaneo, F., Hughes, D.W., \& Kim, E-J. 1996, \prl, 76, 2057
\bibitem[Childress \& Gilbert(1995)]{526}
	Childress, F. \& Gilbert, A.D. 1995, {\em Stretch, Twist, Fold: The Fast Dynamo}, Springer
\bibitem[Courvoisier, Hughes, \& Tobias(2006)]{528}
	Couvoisier, A., Hughes, D.W. \& Tobias, S.M. 2006, PRL, 96, 034503
\bibitem[Galloway \& Proctor(1992)]{GAP1992}
        Galloway, D.~J. \& Proctor, M.~R.~E.\ 1992, \nat, 356, 691
\bibitem[Gilbert(2002)]{532}
	Gilbert, A.D. 2002, Phys. Rev. D, 166, 167
\bibitem[Goldreich \& Sridhar(1995)]{534}
	Goldreich, P. \& Sridhar, S. 1995, \apj, 438, 763
\bibitem[Goldreich \& Sridhar(1997)]{536}
        Goldreich, P. \& Sridhar, S. 1997, \apj, 485, 680
\bibitem[Haugen, Brandenburg, \&Mee(2004)]{}
	Haugen, N.E., Brandenburg, A., \& Mee, A.J. 2004, \mnras, 353, 947
\bibitem[Heitsch et al.(2007)]{538}
        Heitsch, F., Slyz, A.D., Devriendt, J.E.G., Hartmann, L.W., Burkert, A.\ 2007, ApJ, 665, 445
\bibitem[Heitsch \& Zweibel(2003a)]{540}
        Heitsch, F. \& Zweibel, E.G.\ 2003, \apj, 583, 229
\bibitem[Heitsch \& Zweibel(2003b)]{542}
        Heitsch, F. \& Zweibel, E.G.\ 2003, \apj, 590, 291
\bibitem[Heitsch et al.(2004)]{544}
        Heitsch, F., Zweibel, E.G., Slyz, A.D., Devriendt, J.E.G.\ 2004, \apj, 603, 165
\bibitem[Kim(1997)]{546}
	Kim, E-J. 1997, \apj, 477, 183
\bibitem[Klapper \& Young(1995)]{548}
	Klapper, I.M. \& Young, L.S.  1995, Comm Math Phys 173, 623
\bibitem[Kulsrud \& Anderson(1992)]{}
	Kulsrud, R.M. \& Anderson, S.W. 1992, \apj, 396, 606
\bibitem[Kulsrud \& Zweibel(2008)]{550}
        Kulsrud, R.M. \& Zweibel, E.G.\ 2008, astro-ph/0707.2783
\bibitem[Moffatt(1978)]{MOF1978}
	Moffatt, H.K. \ 1978, in {\em Magnetic field generation in electrically conducting
	fluids}, Cambridge Univ. Press, Cambridge
\bibitem[Ott(1998)]{555}
	Ott, E. 1998, Phys. Pl. 5, 1636
\bibitem[Proctor \& Zweibel (1992)]{}
	Proctor, M.R.E. \& Zweibel, E.G. 1992, GAFD, 64, 145
\bibitem[Schekochihin et al.(2004)]{557}
	Schekochihin, A.A., Cowley, S.C., Taylor, S.F., Maron, J.L., \& McWilliams, J.C. 2004, \apj, 612, 276
\bibitem[Schekochihin et al.(2007)]{559}
        Schekochihin, A.A., Isakov, A.B., Cowley, S.C., McWilliams, J.C., Proctor, M.R.E., \& Yousef, T.A.  2007, New J Phys, 9, 300
\bibitem[Shu(1983)]{}
	Shu, F.H. 1983, \apj, 273, 202
\bibitem[Sridhar \& Goldreich(1994)]{561}
	Sridhar, S. \& Goldreich, P. 1994, \apj, 432, 612
\bibitem[Subramanian(1997)]{}
	Subramanian, K. 1997, arXiv:astro-ph/9708216
\bibitem[Subramanian(1999)]{}
	Subramanian, K. 1999, PRL, 83, 2957
\bibitem[Tanner \& Hughes(2003)]{563}
        Tanner, S.E.M. \& Hughes, D.W.\ 2003, \apj, 586, 685
\bibitem[Tang \& Xu(2000)]{TAX2000}
        Tang, H.~Z. \& Xu, K.\ 2000, J. Comp. Phys., 165, 69
\bibitem[Xu(1999)]{XUK1999}
        Xu, K.\ 1999, J. Comp. Phys., 153, 334
\bibitem[Zweibel (1988)]{}
	Zweibel, E.G. 1988, \apj, 329, 384
\end{thebibliography}
\end{document}